\documentclass[notitlepage,secnumroman,amssymb, nofootinbib,reprint, superscriptaddress, aps]{revtex4-1}
\usepackage{amsmath,amsfonts,physics,relsize}
\usepackage[capitalise]{cleveref}
\usepackage[caption=false]{subfig}
\usepackage{graphicx,color}
\usepackage[normalem]{ulem}

\newcommand{\iu}{\mathrm{i}}
\newcommand{\e}{\mathrm{e}}
\newcommand{\QUAD}{\quad\quad\quad\quad\quad}

\newcommand{\argl}{\left(x-\tfrac{1}{2}y\right)}
\newcommand{\argr}{\left(x+\tfrac{1}{2}y\right)}
\newcommand{\ff}{\mathfrak{f}}
\newcommand{\Vh}{\mathcal{V}_H}
\newcommand{\Th}{\mathcal{T}_H}
\newcommand{\inner}[2]{\langle #1,#2\rangle}

\newcommand{\TV}{\widetilde{V}}
\newcommand{\TT}{\widetilde{T}}

\crefname{appsec}{Appendix}{Appendices}
\graphicspath{{figures/}}

\begin{document}

\title{Quantum fluctuations inhibit symmetry breaking in the HMF model  }%

\author{Ryan Plestid}%
\email{plestird@mcmaster.ca}
\affiliation{\mbox{Department of Physics \& Astronomy, McMaster University, 1280 Main St. W., Hamilton, Ontario, Canada}}
\affiliation{Perimeter Institute for Theoretical Physics, 31 Caroline St. N., Waterloo, Ontario, Canada}  
\author{James Lambert}
\email{lambej3@mcmaster.ca}
\affiliation{\mbox{Department of Physics \& Astronomy, McMaster University, 1280 Main St. W., Hamilton, Ontario, Canada}}
\date{\today}%
\begin{abstract}
\setlength{\parskip}{12pt}
\setlength{\parindent}{0pt}

It is widely believed that  mean-field  theory is exact for a wide-range of classical long-range interacting systems. Is this also true once quantum fluctuations have been accounted for? As a test case we study  the Hamiltonian Mean Field (HMF) model for a system of indistinguishable bosons which is predicted (according to mean-field theory) to undergo a second-order  quantum  phase transition at zero temperature. The ordered phase is characterized  by a spontaneously broken  $O(2)$ symmetry, which,  despite occurring in a one-dimensional model,  is not ruled out by the Mermin-Wagner theorem due to the presence of long-range interactions. Nevertheless, a spontaneously  broken symmetry implies gapless Goldstone modes whose large fluctuations can restore broken symmetries.  In this work, we study the influence of quantum fluctuations by projecting the Hamiltonian onto the continuous subspace of symmetry breaking mean-field states.  
We find that the energetic cost of gradients in the center of mass wavefunction inhibit the breaking of the $O(2)$ symmetry, but that the energetic cost is very small --- scaling as $\order{1/N^2}$. Nevertheless, for any finite $N$, no matter how large, this implies that the ground state has a restored $O(2)$ symmetry. Implications for the finite temperature phases, and classical limit, of the HMF model are discussed. 

\end{abstract}
\maketitle

Systems with long-range interactions lie beyond the scope of traditional statistical mechanics \cite{Dauxois2002a,Campa2009,Levin2014}. They can exhibit ensemble inequivalence \cite{Barre01,Dauxois2002a,Casetti2007,Campa2009,Kastner2010}, divergent relaxation time scales (scaling as $t\gtrsim \log N$ for $N$ particles) \cite{Spohn1992}, and non-ergodic dynamics that lead to late-time states that disagree with the microcanonical ensemble \cite{Goldstein1969,Lecar1972,Barre2002,Barre2002a,Yamaguchi2004,Yamaguchi2008,Joyce2011,Yamaguchi2015} (preferring instead Lynden-Bell \cite{Lynden-Bell1967}, or core-halo statistics \cite{Pakter2011,Teles2011,Levin2014}). While these features were first appreciated  in the context of self-gravitating  systems \cite{Lynden-Bell1967,Binney2011}, it has become  increasingly clear that  peculiarities  of gravitational systems (such as negative specific heat \cite{Lynden-Bell1999} and the gravothermal heat catastrophe \cite{Lynden-Bell1968}) are special cases of a broader statistical theory of long-range interacting systems \cite{Dauxois2002a,Campa2009,Levin2014}. 

One important feature of long-range interactions is that fluctuations can be suppressed to such a degree that continuous symmetries can be spontaneously broken even in one-dimensional systems \cite{Dyson1969,Fisher1972,Bruno2001,Dauxois2002a,Campa2009,Mueller2012,Maghrebi2017}. This can be understood in the context of lattice models  by considering the coordination number of each lattice site. Long-range interactions lead to large coordination numbers, which is equivalent to considering the lattice in some effective dimension $d_\text{eff}>d$. Given that fluctuations are well known to be suppressed in high dimensional systems it is not surprising that long-range interactions can achieve the same effect. Mathematically the presence of long-range interactions invalidates the  Mermin-Wagner theorem \cite{Mermin1966}, and its inapplicability is what allows for spontaneous symmetry breaking in a low-dimensional long-range interacting system \cite{Bruno2001,Campa2009,Maghrebi2017}.

There is an extensive literature concerning the validity of mean-field theory for long-range interacting systems. For instance, in the classical literature, it has been rigorously proven \cite{Braun1977} (i.e.\ with bounded error) that a long-range interacting  system's  exact dynamics are well approximated by a mean-field collisionless-Boltzmann (i.e.\ Vlasov) equation (see e.g.\ \cite{Levin2014} or \cite{Spohn1992}). This approximation is valid on time scales of order $t\lesssim \order{\log  N}$ . Therefore,  in the $N\rightarrow \infty$ limit, it is often said that the collisionless-Boltzmann equation (i.e. mean-field theory) is exact \cite{Braun1977,Spohn1992,Levin2014}. Similar claims exist for equilibrium  physics. For instance,  Lieb was able to rigorously bound the difference between a self-gravitating bosonic star's ground state energy and its Hartree energy, and showed that this difference vanishes in the thermodynamic limit \cite{Lieb1987}. Similarly, it is well known that  long-range interacting spin models  have mean-field critical  exponents \cite{Fisher1972}, and it was  conjectured that their free-energy is also identical to that derived via  a  mean-field (meaning  all-to-all  interacting)  model \cite{Cannas2000}.  Subsequent studies supported  this  idea for long-range interacting spin systems \cite{Campa2000,Tamarit2000,Barre2005,Campa2010}, while more recent work has revealed disagreements between the all-to-all and power-law decaying models in a limited region of parameter space \cite{Mori2010,Mori2011,Mori2013}.

The success of all-to-all models in describing the thermodynamics of long-range interacting systems has led  to  them being an essential building  block upon which modern statistical theories of long-range interacting systems are built. Examples include (for a review see \cite{Campa2009}) the Emery-Blume-Griffiths model \cite{Barre01}, the mean-field $\phi^4$ model \cite{Hahn2006,Campa2006,Campa2007}, and the Hamiltonian Mean Field (HMF) model \cite{Antoni1995}. 

Since its  proposal in 1995 \cite{Antoni1995}  the HMF model has been perhaps the most influential toy model in the long-range interacting community. Originally proposed as a simplified model of self-gravitating systems,  it  has emerged as a paradigmatic starting point, and tool, for understanding generic features of long-range interacting systems. While all-to-all (i.e. mean field) models were motivated above by appealing  to equilibrium  physics, they turn out to also capture dynamical  behavior
.  The HMF model can  be used as a tool for understanding chaos \cite{ Latora1999,Ginelli2011,Manos2011,Filho2018}, violent relaxation \cite{Barre2002,Barre2006a,Pakter2011,Ettoumi2011,Plestid2018,Giachetti2019}, core-halo  statistics \cite{Pakter2011,Teles2011,Levin2014}, and other  quasi-stationary states \cite{Barre2002,Barre2002a, Yamaguchi2004,Barre2006a} in  long-range interacting systems. The HMF model exhibits a second  order phase transition associated with the spontaneous breaking of a continuous [$O(2)$] symmetry \cite{Antoni1995,Dauxois2002a}. The HMF model's canonical partition function can be calculated exactly in the classical limit, and exhibits ensemble equivalence with the microcanonical ensemble.

The HMF model describes particles of unit mass, on a circle of unit radius, interacting via a pairwise cosine potential. When quantized for $N$ indistinguishable bosons, the HMF  model is defined by the Hamiltonian \cite{Chavanis2011}
\begin{equation}\label{ham-hmf}
    \hat{H}_\textrm{HMF}=\sum_i \frac{\chi^2}{2} \pdv[2]{\theta_i} - \frac1N\sum_{i<j} \cos(\theta_i-\theta_j)~.
\end{equation}
Here $\chi$ is a dimensionless Planck's constant\footnote{For a ring of radius $R$, with particles of mass, $m$, and a prefactor of $\epsilon$ multiplying the cosine interaction $\chi=\hbar/\sqrt{m R^2 \epsilon}$.} and we have chosen the case of the attractive HMF model as indicated by the negative sign of the potential.  The $\frac1N$ scaling in front of the cosine interaction, known as the Kac prescription \cite{Kac1963}, preserves  extensivity of  the Hamiltonian, and is a consequence of  the peculiar thermodynamic limit for long-range interacting systems ($N\rightarrow \infty$ with system size, and $\chi$ held fixed \cite{Barre2006,Campa2009,Chavanis2011}). \Cref{ham-hmf} can also be interpreted as a describing a lattice of $O(2)$ quantum rotors interacting with one another via all-to-all interactions with $\theta_i$ labeling the angle of each rotor on the lattice. 

As mentioned above, classically (in the limit $\chi\rightarrow 0$) the model undergoes a thermal clustering transition \cite{Dauxois2002a} characterized by the order parameter 
 \begin{equation}\label{mag-def}
    \vb{M}= \langle \cos\theta \rangle \hat{\vb{x}} + \langle \sin\theta \rangle \hat{\vb{y}} ~,
 \end{equation}
which, for $\vb{M}\neq 0$, implies a spontaneously  broken $O(2)$ symmetry.  This transition is second-order, and is driven by thermal fluctuations. At high temperatures, $T> T_c$, the system is homogeneous, and $\vb{M}=0$. For low temperatures, $T<T_c$, the system spontaneously breaks its underlying $O(2)$ symmetry.  

In addition to mimicking certain dynamical features of self-gravitating bosons \cite{Chavanis2011} \cref{ham-hmf} is also closely related to a handful of quantum systems that can be realized in the lab. For instance, cold atoms loaded into optical cavities can realize the generalized HMF model, which is identical to \cref{ham-hmf} up to terms of the form $\sum_{i<j} \cos[\theta_i-\theta_j]$ \cite{Schutz2014,Keller2018}. If, in the rotor interpretation of the model, the sum in  \cref{ham-hmf} is restricted to be nearest neighbor rotors then it can be shown that this nearest-neighbor quantum rotor model offers a low energy description of bosons in an optical lattice \cite{Sachdev2000Book} (i.e.\ a coupled set of Bose-Josephson junctions). Likewise a spin-$S$ Heisenberg ladder  with anti-ferromagnetic coupling can realize the $O(3)$ nearest neighbor quantum rotor model \cite{Sachdev2000Book}. We would expect an  infinite-range rotor model such as \cref{ham-hmf} to reproduce the physics of rotors with  long-range (i.e. polynomially decaying $1/|r_i -r_j|^{\alpha}$ with $\alpha <1$) couplings as can be engineered in trapped ion systems \cite{Kim2009,Zeiher2017,Islam2013}. 

 Chavanis  undertook the first study of the  model's bosonic \cite{Chavanis2011} (and fermionic \cite{Chavanis2011a}) equilibrium phase diagram. Using a Hartree ansatz, one  finds that for $\chi<\sqrt{2}$ the lowest Hartree-energy state also spontaneously breaks the $O(2)$ symmetry, becoming a  delta-function in the limit that $\chi\rightarrow  0$. For $\chi>\sqrt{2}$ it is found that  the gradient energy for a single-particle wavefunction is no longer compensated for by the gain in interaction  energy; this leads to a homogeneous ground state. Both of these two behaviors connect smoothly with the model's limiting cases: For  $\chi\rightarrow 0$  this  agrees with  the $T\rightarrow 0$ prediction of the classical  HMF model. For $\chi\rightarrow\infty$  we recover  an ideal Bose gas in a finite volume,  the  ground state of which is indeed a homogeneous product state. 
 
 Recently, the HMF model has been considered as a quantum dynamical system. The  quantum analogue of certain classical behaviors, such as violent relaxation, and the formation of quasi stationary states has been studied \cite{Plestid2018}. Interestingly, classical instabilities related to the formation  of bi-clusters \cite{Barre2002,Barre2002a} have been found to be stabilized by  quantum (kinetic) pressure \cite{Plestid2018}. The HMF model's Gross-Pitaevskii equation has also been found to admit exactly  solvable solitary wave solutions \cite{Plestid2018a}. In fact, the Hartree states considered by Chavanis \cite{Chavanis2011} may be considered as a special case of these solutions.  
 
 In this paper we make use of the exact solutions of \cite{Plestid2018a} to systematically study whether quantum effects beyond mean-field theory  can  modify the HMF model's symmetry breaking pattern at zero temperature.  In  particular,  the mean-field (i.e.  Hartree) prediction of a spontaneously broken [$O(2)$] symmetry suggests a highly degenerate ground state;  if there is one ground state $\ket{\Theta}$ with its center  of mass at $\Theta$, then there must be continuous manifold of such states  $\{\ket{\Theta'}\}$ with  $\Theta'\in[-\pi,\pi)$. This is reminiscent, for instance, of spinor Bose-Einstein condensates, whose exact ground state is a continuous quantum superposition of mean-field solutions \cite{Ho2000,Castin2000}; we term these states \emph{continuous cat states} (CCS). In our example, such states would correspond  to fluctuations of the center of mass, or, equivalently of a low-lying Goldstone excitation related to the broken $O(2)$ symmetry.
 
 We focus on computing matrix  elements  of the Hamiltonian between different Hartree states, $\bra{\Theta}\hat{H}_\text{HMF}\ket{\Theta'}$. Translational invariance, and parity,  ensures that these matrix elements can depend only on the  difference $|\Theta-\Theta'|$.  Then, since the Hartree states tend towards delta functions, we can expect a delta-expansion (in terms of  derivatives of the Dirac-delta  function)  to provide a good approximation of their behavior.  Projecting the Hamiltonian  onto this subset of states and using this expansion  we may then infer whether  or  not  quantum  fluctuations of  the center of mass raise, or lower, the energy. 
 
 Viewing the HMF model as archetypal of long-range interacting systems, it is natural to study how the model's  phase  diagram is modified by quantum effects. Mapping  out the phase diagram  for  the HMF model in the $\chi-T$ plane is a natural,  and important, addition to the cannon of literature surrounding the HMF model. In this paper we take the first step towards this goal by studying the role of quantum fluctuations at zero temperature. 
 

 The rest of the paper is dedicated to calculating the energetic cost (or profit) of center of mass fluctuations as sketched above. In \cref{mft} we review the Hartree analysis for the HMF model \cite{Chavanis2011,Plestid2018a} which will serve as a starting point for our analysis. In \cref{com-fluc} we calculate matrix elements of the Hamiltonian between different CCS. In \cref{large-n} we develop a large-$N$ asymptotic series for the energy of a given CCS. Then, in \cref{strong-coup} we obtain explicit expressions for the energy at leading order in $\chi$; this allows us to determine the symmetry breaking properties of the ground state. 
 Finally, in \cref{conclusions} we summarize our results and suggest future directions for the quantum HMF model.




\section{Mean Field Theory \label{mft}}
Mean-field theory for the bosonic HMF model  at zero temperature is equivalent to a product-state ansatz for the ground state. Taking $\ket{\Psi}=\bigotimes\ket{\psi}$, with $\ket{\psi}$ a single particle state leads to an energy functional $\mathcal{E}[\psi]=\bra{\Psi}\hat{H}_\text{HMF} \ket{\Psi}$.  Minimizing this energy with respect to the single particle wavefunctions,  $\delta \mathcal{E}/\delta\psi=0$ then leads to a self-consistent eigenvalue problem \cite{Plestid2018a}
\begin{equation}\label{GGPE}
    -\frac12\partial_\theta^2 \psi_H + M\cos\theta\psi_H = \mu \psi_H 
\end{equation}
where $\mu$ is the chemical potential, and $M$ is the aforementioned order parameter of \cref{mag-def}; in the Hartree theory, $M$ must be determined self-consistently. \Cref{GGPE} is exactly soluble, and its solutions can be expressed in terms of Mathieu functions \cite{Plestid2018a,Olver2010}
\begin{equation}\label{psiH-mathieu}
    \psi_H(\theta)=  \frac1{\sqrt\pi} \mathrm{ce}_0\qty(\tfrac{\theta-\pi}{2};q[\chi])~,
\end{equation}
where $q(\chi)$ is the depth-parameter of the Mathieu equation \cite{Olver2010}, whose dependence on $\chi$ can be  determined by solving the  self-consistency condition
\begin{equation}
    q=\frac{4M}{\chi^2}
\end{equation}
In this context, the magnetization may be thought of as a function of $q$, and is defined via  the integral 
\begin{equation}
    M(q)=\frac{1}{\pi} \int_{-\pi}^\pi \qty[\mathrm{ce}_0\qty(\tfrac{\theta-\pi}{2};q)]^2\cos\theta~.
\end{equation}
Solving \cref{GGPE}, one finds that for $\chi>\sqrt{2}$  the magnetization vanishes,  $M=0$, and that the lowest energy wavefunction  is homogeneous i.e. $\psi_H=1/\sqrt{2\pi}$ \cite{Chavanis2011}. Furthermore, a Bogoliubov theory of fluctuations about this ground state can be constructed, and it can be easily checked that the quantum depletion of the ground state  $\sum_{k\neq 0} \langle a^\dagger_k a_k\rangle_{T=0}$  (with $a_k$ the atomic ladder operator) is finite, being given by $\sum_{k\neq 0}  \sinh^2\theta_k$ with $\sinh^2\theta_k= \frac12\qty(\sqrt{1-2\delta_{k,\pm1}/\chi^2}-1)$. 

For $\chi<\sqrt{2}$ one finds instead that $M\neq0$ and the ground state wavefunction begins to acquire non-zero curvature, with the explicit wave-function being give by \cref{psiH-mathieu}. The transition between the spatially homogeneous ground state and  the spatially localized ground state can be viewed as a quantum phase transition  associated with the spontaneous breaking of translational invariance; the transition  is predicted to be second order. 

 This simplified analysis then predicts that there is a degenerate manifold of ground states, given by  $\ket{\Theta;N}=\bigotimes \ket{\psi_H; \Theta}$ where $\Theta$ labels the wavefunction's center of mass (COM), such that  $\psi_H(\theta-\Theta)=\braket{\theta}{\psi_H; \Theta}$ is peaked at $\theta=\Theta$. For this kind of mean-field  analysis  to be self-consistent, however, we require that quantum fluctuations of the COM are small \emph{a posteori}. 
 Because the clustered phase is characterized by a spontaneously broken continuous symmetry, we must then consider  fluctuations of gapless excitations corresponding to the shift symmetry $\Theta \rightarrow \Theta + \Delta \Theta$. 


\section{Center of mass fluctuations \label{com-fluc}}
We can study the importance of COM fluctuations by considering a CCS 
\begin{equation}
\ket{\ff; N}= \int \dd \Theta~ \ff(\Theta) \ket{\Theta; N}
\end{equation}
where $\ff$ is the COM  wavefunction, such  that $\ket{\ff;N}$ is  a  coherent  superposition of product states centered about $\Theta$. The product states 
\begin{equation}
    \ket{\Theta; N}= \bigotimes_{i=1}^N \ket{\psi_H;\Theta }
\end{equation}
are composed of single particle wavefunctions, centered  at  $\Theta$, $\psi_H(\theta-\Theta)=\braket{\theta}{\psi_H;\Theta}$, that minimize the Hartree (i.e. mean-field) energy. 

The case of $\ff\propto \delta(\Theta)$ corresponds to a Hartree state (localized about a single COM) whereas if $\ff(\Theta)$ is independent of $\Theta$ then this state has a restored translational invariance. To test whether or not quantum fluctuations restore translational symmetry we can compute  the average energy of a CCS. We are therefore interested in minimizing the energy-per-particle 
\begin{equation}
E[\ff]=\frac{1}{N}\bra{\ff; N} \hat{H}_\text{HMF} \ket{\ff; N} ~.
\end{equation} 
Because of the system's translational invariance, we can guarantee that the resulting functional can be diagonalized in momentum space
\begin{equation}
E[\ff]= \sum_k  \hat{E}(k) |\hat{\ff}(k)|^2, 
\end{equation} 
where $\ff(\Theta)= \sum_k \e^{\iu k \Theta} \hat{\ff}(k)/\sqrt{2\pi}$. Studying the variational problem $\delta E/\delta \ff=0$ is equivalent to minimizing $\hat{E}(k)\abs{\ff(k)}^2$ subject to the constraint that $\braket{\ff;N}{\ff; N}=1$; we can therefore conclude without any loss of generality that the minimum energy COM wavefunction will be of the form 
$\ff_k(\Theta)= \e^{\iu k \Theta}$ with $k\in \mathbb{Z}$.


It will be useful to introduce the functions 
\begin{equation}\label{thy}
\mathcal{T}_H(y)= \frac{\chi^2}{2}\int \dd \theta \partial\psi_H^*(\theta-\tfrac12 y)\partial\psi_H(\theta+\tfrac12 y)
\end{equation}
and 
\begin{equation}\label{vhy}
\begin{split}
\mathcal{V}_H(y)&=\frac{1}{2}\int \dd \theta \dd \theta' \psi_H^*(\theta-\tfrac12 y)\psi^*_H(\theta'-\tfrac12 y) \\
&~\quad\times \cos(\theta-\theta')\psi_H(\theta+\tfrac12 y)\psi_H(\theta'+\tfrac12 y)~,
\end{split}
\end{equation}
with $y=\Theta-\Theta'$.
Throughout our analysis we will find that the Hamiltonian's matrix elements between two Hartree states $\bra{\Theta_2;N} \hat{H}\ket{\Theta_1;N}$  can be expressed in terms of derivatives of the above functions evaluated at zero separation. With this in mind we introduce the following notation
\begin{equation}
    \Vh^{(n)}= \eval{\partial^n_y \Vh(y)}_{y=0}~\text{and}~~\Th^{(n)}=\eval{\partial^n_y \Th(y)}_{y=0}~.
\end{equation}
\vspace{-6pt}

Explicitly, the functions $\Vh(y)$ and $\Th(y)$ are related to the the matrix elements of the kinetic, $\hat{T}=\sum_i \tfrac{1}{2m}\hat{p}_i^2$,  and potential, $\hat{V}=-\tfrac1N \sum_{ij} \cos(\hat{\theta}_i -\hat{\theta}_j)$,  operators  via 
\begin{equation}
   \frac1N \bra{\Theta_1 ; N} \hat{T} \ket{\Theta_2;N} =\mathcal{T}_H(y) O(y;N-1)
\end{equation}
\begin{equation}
   \frac1N\bra{\Theta_1 ; N} \hat{V} \ket{\Theta_2;N} = -\mathcal{V}_H(y)\qty[\frac{N(N-1)}{N^2}] O(y;N-2)
\end{equation}
where we define the overlap 
\begin{equation}
O(y;\aleph)=\braket{\Theta_1 ; \aleph}{\Theta_2 ; \aleph}=
\qty[\braket{\psi_H;\Theta_1}{\psi_H;\Theta_2}]^\aleph~,
\end{equation}
(with $\aleph=N,~N-1,~\text{or},~N-2$), in terms of the coordinate difference $y=\Theta_1-\Theta_2$.  Introducing the COM coordinate $x=\tfrac12(\Theta_1+\Theta_2)$ we can write
\begin{widetext}
\begin{equation}\label{e-exact}
E[\ff]= \int_{-\pi}^{\pi}\int_{-\pi}^{\pi} \dd x \dd y~ \ff^*\argl~\ff\argr~\bigg[ \mathcal{T}_H(y) O(y;N-1)-\mathcal{V}_H(y)\qty[1-\tfrac1N] O(y;N-2) \bigg] ~,
\end{equation}
from which we can immediately see that 
\begin{equation}\label{ehat-exact}
 \hat{E}(k)=\int_{-\pi}^{\pi}\dd y~ \e^{-\iu k y} \bigg[ \mathcal{T}_H(y) O(y;N-1)-\mathcal{V}_H(y)\qty[1-\tfrac1N] O(y;N-2) \bigg] ~. 
\end{equation}
\end{widetext}
Note that in both \cref{e-exact,ehat-exact} all of the derivatives from the full many body Hamiltonian \cref{ham-hmf} appear in the functions $\Th(y)$ and $\Vh(y)$ and they do not act directly on the COM wavefunction. 

\section{Large-$N$ Expansion \label{large-n}}
We are ultimately interested in the thermodynamic limit ($N\rightarrow \infty$ with $\chi$ held fixed \cite{Barre2006,Chavanis2011}), and in particular whether sub-leading corrections in $1/N$ can modify the symmetry breaking pattern at zero temperature. To study this limit we develop an expansion that relies on the $\aleph$-body overlap, $O(y; \aleph$), being tightly peaked for $\aleph\gg 1$. Because $O(y;\aleph)=\qty[\braket{\psi_H;\Theta_1}{\psi_H;\Theta_2}]^\aleph$ can be written as an exponentiated single particle overlap, this will be true even for moderately peaked single-particle overlaps. In the clustered phase, provided $\chi\lesssim 1$, the overlap between two Hartree states, $\ket{\Theta_1;\aleph}$ and $\ket{\Theta_2;\aleph}$, admits a $\delta$-expansion of the form
\begin{align}\label{overlap-expansion}
O(y;\aleph) &= \frac{1}{\abs{C(\aleph,\chi)}^2}\qty[\delta(y)+ \sum_{p>0} \frac{\mathcal{K}_p(\chi)}{\aleph^p} \delta^{(2p)} (y) ]~.
\end{align}
We use this to develop a systematic expansion in $1/N$ by considering a perturbative expansion of the COM wave-function 
\begin{equation}\label{f-asym-def}
\ff= C(N,\chi)\qty( \ff_0 + \frac1N \ff_1 + \frac1{N^2}\ff_2+...).
\end{equation}
The multiplicative constant $C(N,\chi)$ is chosen such that $\langle \ff_0, \ff_0\rangle = \int \ff_0^*(x) \ff_0(x)\dd x =1$, which ensures that $\braket{\ff;N}{\ff;N}=1$ at leading order\footnote{The  physical state overlap$\braket{\ff}{\ff}$ differs  from the   $L^2$ inner   product, $\langle \ff, \ff \rangle$   at $\order{1/N}$ i.e.  $\braket{\ff}{\ff}  =  \langle \ff, \ff \rangle + \order{1/N}$. }. To maintain this normalization order by order in $1/N$ we impose the following constraints on the COM wavefunction
\begin{align}
  \langle \ff_0 , \ff_0 \rangle &=1\label{normalization-condition1}\\
   2\mathrm{Re} \langle \ff_1, \ff_0\rangle &=\mathcal{K}_1 \langle  \ff_0^{(1)},  \ff_0^{(1)}\rangle\label{normalization-condition2} \\
   \langle \ff_1 , \ff_1 \rangle + 2 \mathrm{Re} \langle \ff_0, \ff_2\rangle &= \mathcal{K}_1 ~2 \mathrm{Re} \langle \ff^{(1)}_1 , \ff^{(1)}_0 \rangle \label{normalization-condition3}\\
   & \QUAD -  \mathcal{K}_2 \langle  \ff^{(2)}_0, \ff^{(2)}_0 \rangle \nonumber
\end{align}
which can be derived using \cref{overlap-expansion} and identity \cref{delta-id-1}.  Note that, as above, the inner product $\langle \ff_i, \ff_j \rangle=\int \dd x \ff_i^*(x)  \ff_j(x)$  is the $\text{L}^2$ inner product, and should not be confused with the state overlap $\braket{\ff}{\ff}$.  

These normalization constraints play an important role in the calculation of the energy as discussed in \cref{large-N-asym}. Due to non-trivial correlations between $\ff_1$ and $\ff_0$,  expanding $\hat{E}(k)$  directly will not tell us how the energy $E[\ff]$ depends on the wavefunction $\ff$. Rather, one must expand $\ff$ and $\hat{E}(k)$ concurrently.

%
\begin{equation}\label{e-hat-exp}
\hat{E}(k)= \frac{1}{|C(N,\chi)|^2}\qty[\hat{E}_0 + \frac{1}{N}\hat{E}_1 +  \frac{1}{N^2}\hat{E}_2 +~ ...~]~,
\end{equation}
where  $\hat{E}_0=E_H$, with $E_H= \Th^{(0)}-\Vh^{(0)}$ the Hartree energy. We find that $\hat{E}$ is given by
\begin{equation}
\label{e-hat-f}
\begin{split}
\hat{E}(k)|\ff(k)|^2&= \hat{E}_0 |\ff_0|^2 + \frac{1}{N}\qty[\hat{E}_1|\ff_0|^2+2\hat{E}_0\mathrm{Re}~\ff_0^*\ff_1]\\
& + \frac{1}{N^2}\qty[ \hat{E}_2 |\ff_0|^2 + \hat{E}_1|\ff_1|^2  +2\hat{E}_0\mathrm{Re}~\ff_0^*\ff_2].
\end{split}
\end{equation}
By using \cref{normalization-condition1,normalization-condition2,normalization-condition3}, the above expression can be simplified such that $E[\ff]=\sum_k \hat{E}(k)|\ff(k)|^2$ can be written as
\begin{equation}
    E[\ff]= E_0 + \frac{E_2}{N^2} \langle \ff_0^{(1)}, \ff_0^{(1)} \rangle + \order{\frac{1}{N^3}}
\end{equation}
or at the same level of accuracy 
\begin{equation}\label{energy-exp}
    E[\ff]= E_0 + \frac{E_2}{N^2} \langle \ff^{(1)}, \ff^{(1)} \rangle + \order{\frac{1}{N^3}}
\end{equation}
\Cref{energy-exp} controls the symmetry breaking in the HMF model. Naively, the term $E_2$ is irrelevant in the thermodynamic limit [being $\order{1}$], however, because the leading order term predicts a degenerate ground state, the small $\order{1/N^2}$ perturbation $\hat{E}_2$ dictates the symmetry breaking pattern of the ground state. The sign of $E_2$ dictates whether in-homogeneity (i.e. non-zero values of $k$) raises or lowers the energy of a CCS, and is consequently indicative of whether or not quantum fluctuations can destroy the localized (magnetized) phase. The full details of our calculation can be found in \cref{large-N-asym}, however for brevity's sake we simply quote the leading order contribution for each quantity
\begin{equation}\label{e0-eval}
 E_0 = E_H + \frac{1}{N}[ \Th^{(2)} - \Vh^{(2)} -\tfrac12 \Th^{(0)}] + \order{\frac{1}{N^2}}~,
\end{equation}
and 
\begin{equation}\begin{split}\label{e2-eval}
E_2= 
[\mathcal{K}_1^2  
    -6\mathcal{K}_2][\Th^{(2)} -\Vh^{(2)}]
     - \mathcal{K}_1[\Th^{(0)}-2\Vh^{(0)}]~.
\end{split}\end{equation}
The fact that gradient corrections vanish at $\order{1/N}$ is a consequence of a cancellation between the $\hat{E}_1|\ff_0|^2$ and $2 \hat{E}_0 \mathrm{Re}~\ff_0^*\ff_1$ in \cref{e-hat-f}. This cancellation is not accidental, and is discussed in greater detail in \cref{cancellations}


\section{Strong Coupling Regime \label{strong-coup}} 

To determine whether these fluctuations can restore translational invariance we can study a point in parameter space deep within the clustered phase $\chi\lesssim 1$ and see if quantum fluctuations can lead to a translationally invariant COM wavefunction (i.e. $\ff =1/\sqrt{2\pi}$). For this to occur $\hat{E}_2$ must be positive such that $k=0$ is energetically preferred. 

Although left implicit until now, the parameters $\mathcal{K}_1(\chi)$, and $\mathcal{K}_2(\chi)$ are themselves functions of $\chi$ as are the derivatives of the CCS energies $\Vh^{(n)}(\chi)$ and $\Th^{(n)}(\chi)$. These functions are determined exactly in terms of integrals \cref{thy,vhy} involving the Hartree ground state $\psi_H(\theta; \chi)$ (whose $\chi$ dependence is determined by \cref{GGPE}).  To test whether quantum fluctuations of the COM can restore the spontaneously broken symmetry it is sufficient to restrict our attention to small but finite values of $\chi$ satisfying $\chi\ll \sqrt{2}$ . 

Both $\mathcal{K}_1$ and $\mathcal{K}_2$ are determined by $O(y; N,\chi)$. As argued in the appendix, for small values of $\chi$  this can be well approximated by (see \cref{overlap-appendix})
\begin{equation}
O(y; \aleph)\approx \qty[\frac{I_0\qty( \sqrt{q}\cos\tfrac{y}{2})}{I_0\qty(\sqrt{q}) } ]^\aleph
\end{equation}
where $I_0(z)$ is the modified Bessel function of the first kind and $q$ is an auxiliary depth parameter related to the mean-field magnetization, $M$, and $\chi$  via $q=\sqrt{4 M/\chi}$ .  We are interested in finding a delta-expansion for $O(y)$ and are thus interested in integrals of the form $\int_{\pi}^\pi O(y) f(y)~\dd y$.  For $1\lesssim y\lesssim \pi$  the overlap is exponentially small [i.e. $\order{\e^{-\sqrt{q}}}$ ]  so we can neglect this contribution to the integral. For moderate values of $y$ we can then use the large argument expansion of the modified Bessel functions $I_0(z)\sim \e^{-z}/\sqrt{2\pi z}$ leading to 
\begin{equation}
 O(y;\aleph)\sim \exp\left\{\aleph\qty[\tfrac{4}{\chi}(1-\tfrac18\chi)\sin^2\tfrac{y}{4} - \tfrac12 \log \cos\tfrac{y}{2} ] \right\} ~.
\end{equation}
Using this exponential form, the integrals we are interested in studying can then be approximated using Watson's Lemma
\begin{equation}
	\int \e^{- \aleph G(y)} f(y) \dd y \sim \sqrt{\frac{2\pi}{\aleph G^{(2)}}} \sum_p \frac{ f^{(2p)}}{(2p)!!\qty[\aleph G^{(2)}]^p},
\end{equation}
where the bracketed superscripts denote the $2p^{\text{th}}$ derivative of the function evaluated at $y=0$. For $O(y;\aleph)$ we have
\begin{equation}
    G^{(2)}= \frac{1}{2 \chi }-\frac{3}{16}.
\end{equation}
We can then read off overall prefactor of \cref{f-asym-def}
\begin{equation}
|C(N, \chi)|^2= \sqrt{\frac{2\pi}{N  G^{(2)} } }=2\sqrt{\frac{\pi   \chi}{N}}\qty[ 1 + \frac{3\chi}{32}  ]
\end{equation}
and the coefficients $\mathcal{K}_1$ and $\mathcal{K}_2$ which are given at next-to-leading order
\begin{equation}
	\mathcal{K}_1\sim \chi+ \frac{3\chi ^2}{8} \quad\quad  \mathcal{K}_2\sim \frac{\chi ^2}{2}+\frac{3\chi ^3}{8}
\end{equation}
Next, using \cref{sips} for the Mathieu functions, we can derive the small-$\chi$ behavior of the CCS-functionals and their derivatives. 
\begin{align}
 \Th^{(0)}&\sim  \frac{\chi}{4}& \Th^{(2)}&\sim- \frac{3}{8}\\
 \Vh^{(0)}&\sim  \frac12  - \frac\chi4 & \Vh^{(2)}&\sim -\frac1{2\chi}+ \frac38 .
\end{align}
Note that we need the sub-leading corrections to $\Th^{(0)}$ and $\Th^{(2)}$ because they are the same order as $\Th^{(0)}$ and $\Th^{(2)}$. 

Including these terms we find that $\order{\chi}$ contribution vanishes, but the $\order{\chi^2}$ contribution does not. We finally arrive at 
\begin{equation}\begin{split}
\hat{E}_2\sim \frac{3\chi^2}{8} + \order{\chi^3}~.
\end{split}\end{equation}
This tell  us that that curvature of the COM wavefunction is energetically unfavorable such that  the system prefers  a  homogeneous CCS over a clumped one. Thus, quantum fluctuations corresponding  to Goldstone modes restore the spontaneously broken translational invariance. The lowest energy state, at all finite values  of $N$  (no matter how large), is given by 
\begin{equation}\label{groundstate}
\ket{\mathrm{GS}}_\text{CCS} =\frac{1}{\sqrt{2\pi}} \int_{-\pi}^{\pi}\dd\Theta\ket{\Theta} + \order{\frac{1}{N} }~.
\end{equation}
As was alluded to earlier, this is reminiscent of spinor Bose-Einstein condensates, whose exact ground state is known to be a CCS that is formally identical to \cref{groundstate}  \cite{Ho2000,Castin2000}.

\section{Discussion and Conclusions \label{conclusions}}

Quantum fluctuations of Goldstone modes can play an important role in determining the zero temperature behavior of a long-range interacting system. 
In the example studied here, properties of the ground state such as its symmetry breaking pattern are left undetermined at the  level of mean-field theory due to a high level of degeneracy  in the energy spectrum. Previous work on Bose stars suggests that this degeneracy is a generic consequence of long-range interactions \cite{Lieb1987}. In  the case of the HMF model, we find that this degeneracy is only lifted at $\order{1/N^2}$ for  any finite $N$ (no matter how large). At zero-temperature this has the striking consequence of leading to a restored $O(2)$ symmetry in the ground state. 

At finite $N$, the system is gapped, $\Delta=3\chi^2/8 N^2$,  with excitations corresponding to departures from a homogeneous COM wavefunction.  In the $N\rightarrow\infty$ limit the system becomes gapless, such that $\ket{\mathrm{GS}}_\text{CCS}$ becomes embedded in a highly degenerate manifold of states, almost all of which break the model's underlying $O(2)$ symmetry. This is reminiscent of the behavior of spin-$1/2$ chains, where a rotationally invariant singlet ground state is separated at finite $N$ from a triplet excitation that breaks rotation invariance. In the $N\rightarrow \infty$ limit the gap closes and the singlet becomes embedded in a degenerate ground state manifold whose low lying excitations are triplets \cite{auerbach2012interacting} in analogy with the clumping excitations in the HMF.

\begin{figure}[h!]
\subfloat[0.4\linewidth][]
{\includegraphics[width=0.48\linewidth]{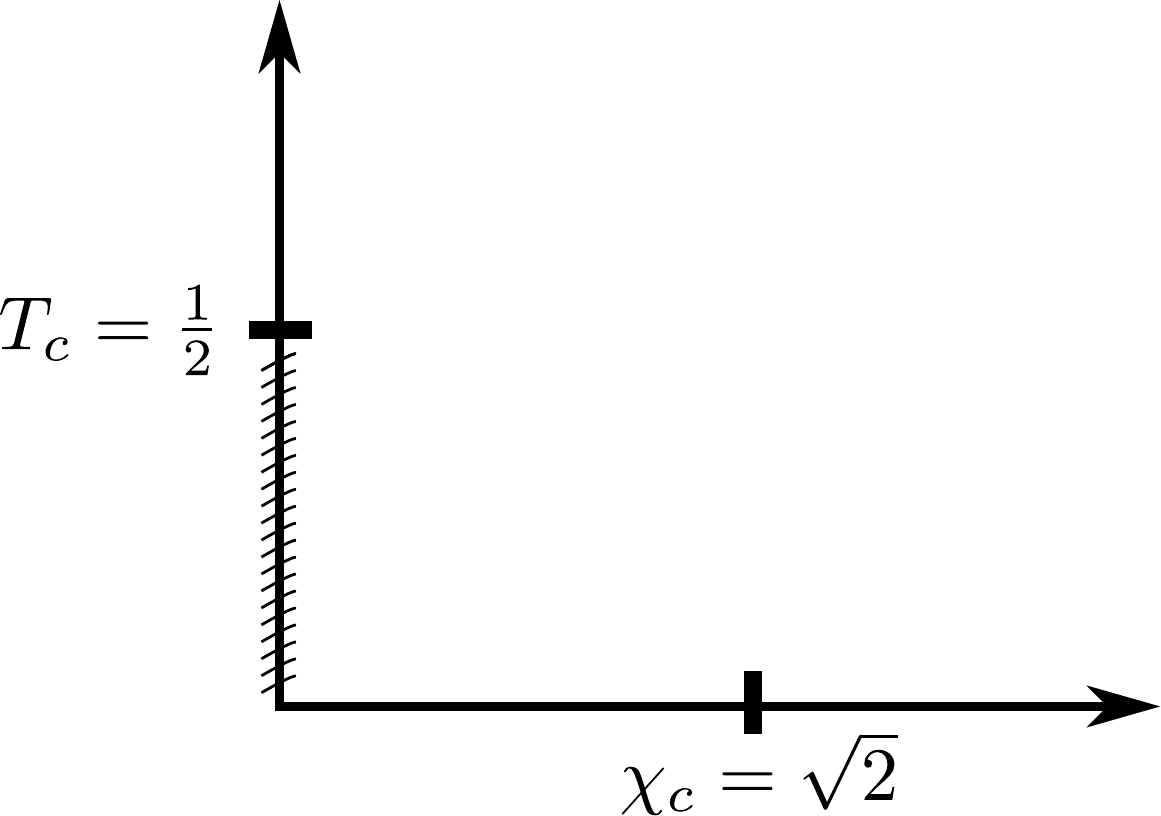}
}%
\subfloat[0.4\linewidth][]
{\includegraphics[width=0.48\linewidth]{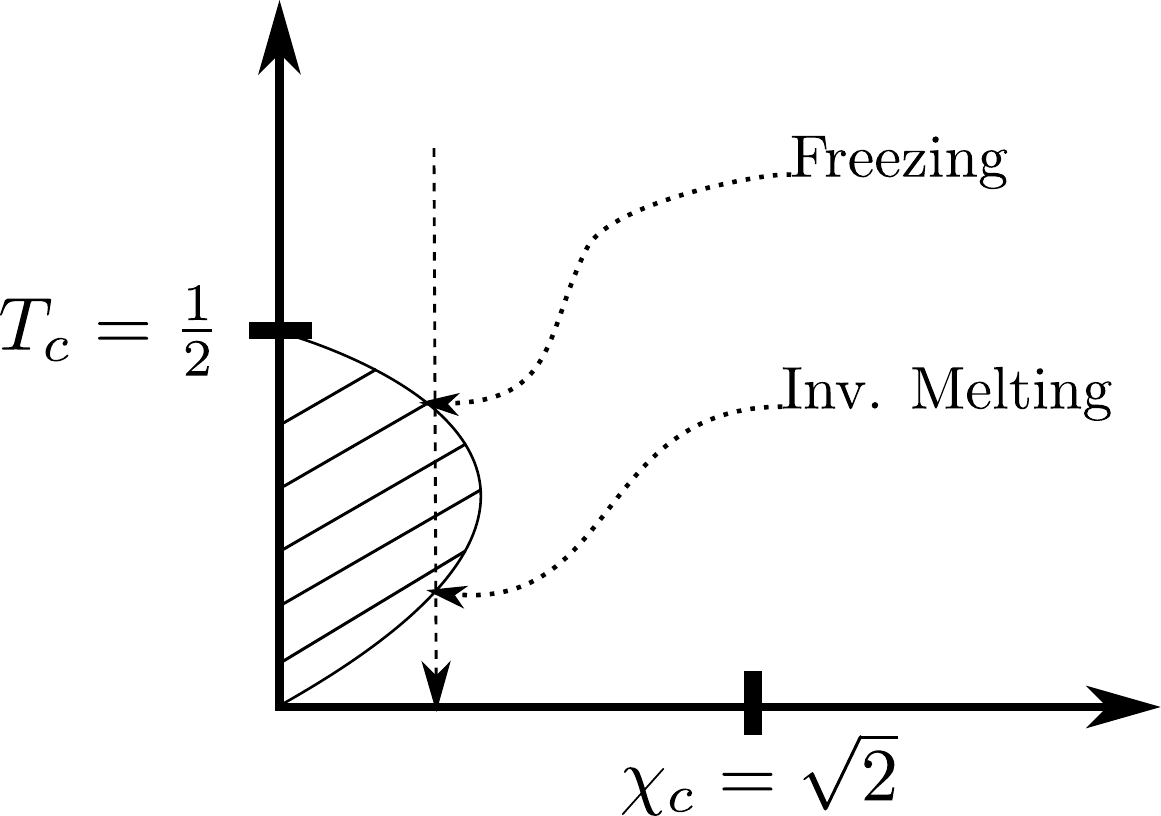}
}
\caption{Two possible resolutions of our result and the exact classical calculation. The limit $\chi\rightarrow0$ could be singular such that symmetry breaking (hashed lines) is  only possible for $\chi=0$ (a).  Alternatively, a re-entrant phase could appear at finite temperature (b). We identify this possibility as analogous to inverse melting, as indicated by the line of decreasing temperature at fixed $\chi$. The parameters corresponding to classical and quantum (mean-field) symmetry breaking are marked with thick black lines. \label{phase-diagram}}
\end{figure}

This discussion  is interesting, because the HMF model's classical partition function can be calculated \emph{exactly} in the $N\rightarrow \infty$ limit, and exhibits a thermally driven second order phase transition \cite{Antoni1995,Campa2009}; at low temperatures the system breaks the $O(2)$ symmetry. Thus, our observation that quantum fluctuations can restore the $O(2)$ symmetry leaves open two logical possibilities that are compatible with the exact classical results \cite{Antoni1995}:
\begin{enumerate}
    \item The limit $\chi\rightarrow0$ is singular, and the classically ordered phase exists only for $\chi$ strictly equal to zero such that for $\chi>0$ quantum fluctuations completely inhibit ordering at all temperatures. 
    \item The HMF model exhibits a re-entrant phase wherein at finite temperature, for small values of $\chi$ the $O(2)$ symmetry is broken. Interpreting the $O(2)$ symmetry as a translational invariance for particles on a ring, this is reminiscent of inverse melting which is known to exist in certain spin models \cite{Schupper2005,Sellitto2006}. 
\end{enumerate}
Schematic phase diagrams for each of these two scenarios  are sketched in \cref{phase-diagram}. The determination of which of these two possibilities is born out by the HMF model is beyond the scope of this paper, however a definitive answer to this question should be attainable via path integral Monte Carlo studies.

The fact that the symmetry of the ground state is protected  by feeble gradient corrections to the energy of the COM wavefunction suggests that the $T\rightarrow 0$ limit is non-trivial. Since deformations of the COM wavefunction should be the lowest energy excitations\footnote{Single particle excitations will have an energy per particle of  $\order{1/N}$, while deformations of the single particle wavefunction $\psi_H\rightarrow\psi_H +\delta \psi$ lead to an energy  per-particle that is  $\order{1}$.}, our analysis suggests that the low-temperature behavior of  the HMF model will be controlled by the parameter $\alpha=\beta \chi^2/N^2$; a cursory examination of this quantity clearly indicates  that the limit of $\beta\rightarrow 0$ (i.e.\ zero temperature) does not commute with $\chi\rightarrow 0$, and, more importantly, $N\rightarrow \infty$. Viewing deformations of the COM wavefunction as low-lying excitations (all of which break the $O(2)$ symmetry) it is conceivable that at finite temperatures it could be entropically favorable to macroscopically excite these degrees of freedom and break the $O(2)$ symmetry. In contrast, one may expect that if feeble quantum fluctuations can inhibit symmetry breaking at zero  temperature, they will continue to be able to do so  at finite temperature.  

In summary we find that quantum fluctuations due to Goldstone modes can substantially alter the symmetry breaking pattern of the HMF model. The energetic cost to excite a non-homogeneous center of mass wavefunction vanishes in the thermodynamic limit, suggesting that finite temperature effects could substantially alter our predictions. While we have provided an analytic study of the HMF model's ground state, our approach is necessarily approximate and we have only included COM fluctuations.  A numerical investigation into both the finite temperature and zero temperature (i.e.\ ground state) properties of the system  is a natural extension of this work, and is the most important next step in the study of the HMF model. 

\section{Acknowledgements}
We would like to thank Dr.\ Sung-Sik Lee for suggestions, discussions, and encouragement. This work was supported by funds from the National Science and Engineering Research Council (NSERC) of Canada.  Support is also acknowledged from the Perimeter     
Institute for Theoretical Physics. Research at the Perimeter Institute is supported by the Government of Canada through the Department of      
Innovation, Science and Economic Development and by the Province of Ontario through the Ministry of Research and Innovation.

\appendix 
 \crefalias{section}{appsec}
  \crefalias{subsection}{appsec}
   \crefalias{subsubsection}{appsec}

\section{Delta function identities}


In \cref{large-N-asym} we frequently encounter integrals  of the form 
\begin{equation}
\int \dd x \dd y \delta^{(2n)}(y) g(x-y/2) f(x+y/2) h(y)~,
\end{equation}
and in this appendix we provide a short derivation of  a useful  identity \cref{delta-id-2}.  We may first, however, study the simpler case of 
\begin{equation}
\int \dd x \dd y \delta^{(2)}(y) g(x-y/2) f(x+y/2).
\end{equation}
In this case we must integrate by parts twice to pull the derivative off of the delta function. This gives 
\begin{equation}
\begin{split}
&=\int \dd x \dd y \delta(y)\frac{1}{4}\qty[g''f + f''g -2 g' f']\\
&=\int \dd x \frac{1}{4}\qty[g''(x)f(x) + f''(x)g(x) -2 g'(x) f'(x)] \\
&= - \int \dd x g'(x) f'(x) .
\end{split}\label{learning-rules}
\end{equation}
This generalizes naturally. If we denote $\partial^{m}[g\argl f\argr  ]=\mathcal{G}_{m}(\alpha,\beta)$ then the generalized identity is 
\begin{equation}\begin{split}
\int \dd x ~\mathcal{G}_{2m}(x,x) &= (-1)^m \int \dd x ~g^{(m)}(x) f^{(m)}(x) \\
&= \langle \bar{g}^{(m)}, f^{(m)}\rangle.\label{G-identity}
\end{split}\end{equation}
Applying this result to a delta-function leads to
\begin{equation}
\begin{split}
&=\int \dd x \dd y~ \delta^{(2n)}(y) g\argl  f\argr   \\
&=\int \dd x \dd y~ \delta(y) \partial_y^{2n}\qty[g\argr  f\argl  ]\\
&= (-1)^n \int \dd x~ g^{(n)}(x) f^{(n)}(x) .
\label{delta-id-1}
\end{split}
\end{equation}
Finally, when including an additional function in the integrand, we simply distribute the derivatives and find 
\begin{equation}
\begin{split}
&=\int \dd x \dd y ~\delta^{(2n)}(y) g\argl  f\argr   h(y) \\
&=\int \dd x \dd y ~ \delta(y)\partial_y^{2n}\qty[g\argr  f\argl h(y)] \\
&=  \int \dd x \dd y~ \delta(y)\sum_m \begin{pmatrix}
2n \\
m \\
\end{pmatrix}  \mathcal{G}_m(\alpha,\beta)\partial_y^{2n-m}h(y)~.
\end{split}
\end{equation} 
Because $h(y)$ is an even function, all of the odd-derivatives vanish leading to
\begin{equation}
\int \dd x \dd y ~\delta(y)\sum_m \begin{pmatrix}
2n \\
2m \\
\end{pmatrix} \mathcal{G}_m(\alpha,\beta) \partial_y^{2(n-m)}h(y) 
\end{equation}
Now we can perform the integration over $y$
\begin{equation}
\begin{split}
&\sum_m \begin{pmatrix}
2n \\
2m \\
\end{pmatrix} \int \dd x \dd y~ \delta(y) \mathcal{G}_{2m}(\alpha,\beta)h^{\qty(2m-2n)}(y) \\
=&\sum_m \begin{pmatrix}
2n \\
2m \\
\end{pmatrix} \qty[\partial_y^{2(n-m)}h(y)]_{y=0}\int \dd x~ \mathcal{G}_{2m}(x,x).
\end{split}
\end{equation}
Now using \cref{G-identity} we arrive at
\begin{equation}\begin{split}\label{delta-id-2}
\int &\dd x \dd y~ \delta^{(2n)}(y) g\argl  f\argr   h(y) \\
&= \sum_m \begin{pmatrix}
2n \\
2m \\
\end{pmatrix} (-1)^m \langle \bar{g}^{(m)}, f^{(m)} \rangle h^{(2n-2m)}_0 ~,
\end{split}
\end{equation}
where $h^{\qty(2n-2m)}_0=\partial_y^{2(n-m)}h(y)\rvert_{y=0}$. In  calculations throughout this paper $f\argr  =\ff_a\argr  $ and $g\argl  =\ff^*_b\argl $ such that $\langle \bar{g}^{(m)}, f^{(m)}\rangle=\langle \ff_b^{(m)},\ff^{(m)}_a\rangle$.


\section{Large-$N$ Asymptotics for the Energy \label{large-N-asym} }
In \cref{e0-eval,e2-eval} we quote  results for ground state energy shift $E_0$, and the COM wavefunction's gradient energy $E_2$. In this appendix we derive these results.   

We begin by considering the kinetic energy  

\begin{equation}\begin{split}
  &T[\ff] = \frac{1}{N}\bra{\ff;N}\hat{T}\ket{\ff;N} \\
  &= \nonumber \int \dd x\dd y\,
  \ff^*\argl\ff\argr\Th(y)O(y;N-1) \\
 \end{split}\end{equation}
 Notice that the overlap has had a particle removed since we are computing the expectation  value of a  single-particle operator.  Because of our COM wavefunction normalization this means we will find an overall prefactor of $|C(N)|^2/|C(N-1)|^2=\sqrt{N/(N-1)}$ . Leading to 
  \begin{align*}
  T[\ff] =& \sqrt{\frac{N}{N-1}}\int\dd x\dd y\, \Th(y)\\
    &\times \sum_{a,b}\frac{1}{N^{a+b}}\ff_a^*\argl\ff_b\argr\\
    &\times\sum_p \frac{\mathcal{K}_p}{N^p}\frac{1}{\left(1-\tfrac{1}{N}\right)^p}\delta^{(2p)}(y).
  \label{}
\end{align*}
where  we have  used $|C(N)/C(N-1)|^2=\sqrt{N/(N-1}$ It is convenient to ignore the prefactor and work with the integral defined above directly. To simplify our analysis we introduce a re-scaled kinetic energy. 
\begin{equation}
  \TT[\ff]=\sqrt{\frac{N}{N-1}}T[\ff] ~,
  \label{}
\end{equation}
such that 
\begin{align}
 \TT[\ff] =& \int\dd x\dd y\, \sum_{a,b}\frac{1}{N^{a+b}}\ff_a^*\argl\ff_b\argr \\
 \nonumber\quad &\times \Th(y) \sum_p \frac{\mathcal{K}_p}{N^p}\frac{1}{\left(1-\tfrac{1}{N}\right)^p}\delta^{(2p)}(y).
  \label{}
\end{align}
If we next consider the potential energy a similar expression may be defined. Starting with
\begin{equation}\begin{split}
   V[\ff]=& \int \dd x \dd y~ \ff^*\argl\ff\argr \\
   &\times \Vh\qty(1-\frac1N) \braket{\Theta_1; N-2}{\Theta_2; N-2}~,
\end{split}\end{equation}
we have 
\begin{equation}\begin{split}
  V [\ff]= \int &\dd x \dd y~ \sum_{a,b}\frac{1}{N^{a+b}}~\ff_a^*(x-\tfrac12y)\ff_b(x+\tfrac12y) \\
    &\times \Vh(y) \times \qty(1-\frac1N)  \abs{\frac{C(N)}{C(N-2)}}^2  \\
    &\times \sum_p \frac{\mathcal{K}_p}{N^p}\frac{1}{\qty(1-\frac2N)^{p}} ~\delta^{(2p)}(y)
 \end{split}\end{equation}
As before we may use $|C(N)/C(N-2)|^2=\sqrt{\frac{N}{N-2}}$ and introduce the function 
\begin{equation}
\TV[\ff]=\sqrt{\frac{N(N-1)^2}{N^2(N-2)}}\TV[\ff]~,
\end{equation}
such that 
\begin{equation}\begin{split}
\TV[\ff] = \int &\dd x \dd y~ \sum_{a,b}\frac{1}{N^{a+b}}~\ff_a^*(x-\tfrac12y)\ff_b(x+\tfrac12y) \\
&\times \Vh(y)  \sum_p \frac{\mathcal{K}_p}{N^p}\frac{1}{\qty(1-\frac2N)^{p}} ~\delta^{(2p)}(y) 
\end{split}\end{equation}
Notice that the expressions for $\TT$ and $\TV$ are nearly identical beyond cosmetic changes such as $\Th\leftrightarrow \Vh$, save for one exception. The sum over $p$ has a factor of $1/(1-m/N)^p$ where $m=1$ for $\TT$ and $m=2$ for $\TV$; this effect enters first at $\order{1/N^2}$ via the term 
\begin{equation}
	\frac{1}{N^2} m \mathcal{K}_1 \delta^{(2)}(y) \qq{$m=1$ or 2}
\end{equation}
At this level of accuracy we therefore have (omitting the explicit arguments of $x\pm \tfrac12 y$ for brevity's sake)
\begin{widetext}
\begin{equation}\begin{split}
  \TT[\ff] = \int \dd x \dd y~ & \qty[ \ff_0^* \ff_0+ \frac1N\qty(\ff_1^* \ff_0 + \ff_0^* \ff_1) + \frac{1}{N^2} \qty( \ff_2^* \ff_0 +  \ff_0^* \ff_2 +  \ff_1^* \ff_1) ]\\
  &\times \Th(y) \qty[  \delta(y) +\frac1N \mathcal{K}_1\delta^{(2)}(y) + \frac1{N^2}\qty( \mathcal{K}_2\delta^{(4)}(y) + \mathcal{K}_1\delta^{(2)}(y) \delta^{(2)}(y) )]
\end{split}\end{equation}
\begin{equation}\begin{split}
  \TV[\ff] = \int \dd x \dd y~ & \qty[ \ff_0^* \ff_0+ \frac1N\qty(\ff_1^* \ff_0 + \ff_0^* \ff_1) + \frac{1}{N^2} \qty( \ff_2^* \ff_0 +  \ff_0^* \ff_2 +  \ff_1^* \ff_1) ]\\
  &\times \Vh(y) \qty[  \delta(y) +\frac1N \mathcal{K}_1\delta^{(2)}(y) + \frac1{N^2}\qty( \mathcal{K}_2\delta^{(4)}(y) + 2 \mathcal{K}_1\delta^{(2)}(y) \delta^{(2)}(y) )]
\end{split}\end{equation}
\end{widetext}
\subsection{Kinetic energy}
At leading order the only contribution to the kinetic energy is given by,
\begin{align}
    \TT_0&= \int\dd x\dd y\, \ff_0^*\ff_0 \delta(y)\Th(y) \nonumber \\
    &= \Th^{(0)}
\end{align}
At next leading order we have,
\begin{align}
    \TT_1 &= \int\dd x\dd y\,\left(\ff_1^*\ff_0 + \ff_0^*\ff_1
    \right)\delta(y) \nonumber \\
    &+ \int\dd x\dd y\,\ff_0^*\ff_0\mathcal{K}_1\delta^{(2)}(y) \nonumber \\
    &=  2 \Re\inner{\ff_0}{\ff_1} + \mathcal{K}_1\sum_{n=0}^1{2\choose 2n}(-1)^n
    \inner{\ff_0^{(n)}}{\ff_0^{(n)}}\Th^{(2-2n)} \nonumber \\
    &=  2 \Re\inner{\ff_0}{\ff_1}\Th^{(0)} -\mathcal{K}_1 \inner{\ff_0^{(1)}}{ \ff_0^{(1)} }\Th^{(0)}  + \mathcal{K}_1\Th^{(2)}
    \nonumber \\
    &= \mathcal{K}_1 \Th^{(2)}\nonumber 
\end{align}
where we have used \cref{delta-id-2}, and in going to the final equality, we have imposed the normalization condition \cref{normalization-condition2}.

At next-to-next-to leading order we have
\begin{align}\label{tt2}
    \TT_2 &= \int \dd x\dd y\, \left(\ff_0^*\ff_2 + \ff_2^*\ff_0 +\ff_1^*\ff_1\right)\delta(y)\Th(y)
    \nonumber \\
    &+ \int\dd x\dd y \left(\ff_0^*\ff_1+\ff_1^*\ff_0\right)\mathcal{K}_1\delta^{(2)}(y)\Th(y)
    \nonumber\\
    &+ \int\dd x\dd y \ff_0^*\ff_0\mathcal{K}_2\delta^{(4)}(y)\Th(y)\nonumber\\
    &+\int\dd x\dd y\ff_0^*
    \ff_0\mathcal{K}_1\delta^{(2)}(y)\Th(y) 
\end{align}
 using  \cref{delta-id-2} and 

\begin{equation}\begin{split}
\TT_2&= 2 \Re\inner{\ff_0}{\ff_2}\Th^{(0)} + \inner{\ff_1}{\ff_1}\Th^{(0)}\\
&+2\mathcal{K}_1\Re\inner{\ff_0}{\ff_1}\Th^{(2)} - 2
    \mathcal{K}_1\Re\inner{\ff_0^{(1)}}{\ff_1^{(1)}}\Th^{(0)}\\
    &+\mathcal{K}_2\Th^{(4)}
    -6\mathcal{K}_2\inner{\ff_0^{(1)}}{\ff_0^{(1)}}\Th^{(2)}+
    \mathcal{K}_2\inner{\ff_0^{(2)}}{\ff_0^{(2)}}\Th^{(0)}\\
    &+\mathcal{K}_1\Th^{(2)} - \mathcal{K}_1
    \inner{\ff_0^{(1)}}{\ff_0^{(1)}}\Th^{(0)}
\end{split}\end{equation}

%
%
%
Summing all of the terms, and   imposing the normalization conditions from \cref{normalization-condition1,normalization-condition2,normalization-condition3}, we find
\begin{equation} \begin{split}
    \TT_2 &= \mathcal{K}_2\Th^{(4)} 
    +\mathcal{K}_1\Th^{(2)} \\
    &+ \qty(\qty[\mathcal{K}_1^2 -6\mathcal{K}_2]\Th^{(2)} -\mathcal{K}_1\Th^{(0)})\inner{\ff_0^{(1)}}{\ff_0^{(1)}} 
\end{split}\end{equation}
In conclusion   we find  
\begin{align}
\TT_0 &= \Th^{(0)}\\
\TT_1  &=\mathcal{K}_1\Th^{(2)}\\
    \TT_2 &= \mathcal{K}_2\Th^{(4)} 
    +\mathcal{K}_1\Th^{(2)} \\
    &~\nonumber+ \qty(\qty[\mathcal{K}_1^2 -6\mathcal{K}_2]\Th^{(2)} -\mathcal{K}_1\Th^{(0)})\inner{\ff_0^{(1)}}{\ff_0^{(1)}} 
\end{align}
Using  $T=(1-\frac1{2N} +\tfrac3{8N^2})\TT$ we then find 
\begin{align}
	T_0 &= \TT_0 \\
	T_1  &=  \TT_1 - \frac12 \TT_0\\
	T_2 &= \TT_2-\frac12  \TT_1  +  \frac38  \TT_0 
\end{align} 
%

\subsection{Potential Energy}
The calculation for $\TV_n$ largely parallels that of $\TT_n$. 
\begin{equation}\begin{split}
    \TV_0&=\int \dd x \dd y~ \ff_0^*\ff_0 \Vh(y) \delta(y)\\
    &= \Vh^{(0)} \langle \ff_0, \ff_0\rangle = \Vh^{(0)}
\end{split}\end{equation}
\begin{equation}\begin{split}
    \TV_1&=\int \dd x \dd y~ \ff_0^* \ff_0 \Vh(y) \mathcal{K}_1 \delta^{(2)}(y) \\
    &\quad\quad\quad\quad+ \qty[\ff_0^* \ff_1 +\ff_1^*\ff_0^*]\Vh(y)\delta(y) \\
    &= \Vh^{(2)} +  2\textrm{Re}\langle \ff_0 ,\ff_1 \rangle \Vh^{(0)} - \mathcal{K}_1 \langle \ff^{(1)}_0, \ff^{(1)}_0\rangle \Vh^{(0)}\\
    &=\Vh^{(2)}
\end{split}\end{equation}
where we have used the COM wavefunction's normalization constraint \cref{normalization-condition2}. 

We then find
\begin{equation}\begin{split}
    \TV_2 =&\int \dd x \dd y ~\ff_0^* \ff_0 \Vh(y) \qty[\mathcal{K}_2 \delta^{(4)}(y) + 2\mathcal{K}_1\delta^{(2)}(y)]\\
 & + \int \dd x \dd y~ \qty[\ff_0^* \ff_1 +\ff_1^*\ff_0^*]\Vh(y) \mathcal{K}_1\delta^{(2)}(y)    \\
 &+ \int \dd x \dd y ~\qty[\ff^*_1  \ff_1  + \ff_0^*\ff_2 + \ff_2^*\ff_0] \Vh(y) \delta(y)~.
    \end{split}
    \end{equation}
 Notice the factor of $2\mathcal{K}_1\delta^{(2)}(y)$ in contrast to the factor of $\mathcal{K}_1\delta^{(2)}(y)$ found in \cref{tt2}.

 As before, we will address each term in the calculation separately, 
\begin{equation}\begin{split}
&= 2 \mathrm{Re}\langle \ff_0 ,\ff_2\rangle \Vh^{(0)} + \langle \ff_1 ,\ff_1\rangle  \Vh^{(0)}  \\
&+2\mathcal{K}_1 \qty[ \Vh^{(2)} - \langle \ff_0^{(1)},\ff_0^{(1)}\rangle \Vh^{(0)}]\\
&+\mathcal{K}_2 \qty[\Vh^{(4)}  - 6 \langle \ff_0^{(1)} ,\ff_0^{(1)}\rangle \Vh^{(2)} + \langle\ff_0^{(2)}, \ff_0^{(2)} \rangle\Vh^{(0)} ]\\
&+ \mathcal{K}_1 \qty[ (2 \mathrm{Re}\langle \ff_0 ,\ff_1 \rangle) \Vh^{(2)} - (2\mathrm{Re}\langle \ff_0^{(1)},\ff_1^{(1)}\rangle )\Vh^{(0)}]
\end{split}\end{equation}

Adding all of these terms together, and making use of the normalization conditions \cref{normalization-condition1,normalization-condition2,normalization-condition3} we find
\begin{equation}\begin{split}
             \TV_2&=\mathcal{K}_2\Vh^{(4)} + 2 \mathcal{K}_1\Vh^{(2)} \\
             &  + \qty(\qty[\mathcal{K}_1^2-6\mathcal{K}_2]\Th^{(2)} - \mathcal{K}_1\Th^{(0)})\langle\ff_0^{(1)},\ff_0^{(1)}\rangle
\end{split}\end{equation}
This leads finally to 
\begin{align}
 \TV_0=&~\Vh^{(0)}\\
 \TV_1=& ~\mathcal{K}_1\Vh^{(2)}\\
 \TV_2 =&~\mathcal{K}_2 \Vh^{(4)} +  2\mathcal{K}_1\Vh^{(2)}\\
 &\nonumber+ \qty(\qty[\mathcal{K}_1^2 -6\mathcal{K}_2 ] \Vh^{(2)}  -2 \mathcal{K}_1\Vh^{(0)} )\langle \ff^{(1)}_0, \ff^{(1)}_0\rangle  .
\end{align}
Lastly we can use the formula $V=(1+\tfrac1{N^2}) \TV + \order{1/N^3}$ to find 
  \begin{equation}
    V_0 = \TV_0\quad\quad  V_1=\TV_1\quad\quad V_2=\TV_2 + \TV_0~.
\end{equation}	
\vspace{6pt}

\subsection{Total Energy} 

Recall that $E[\ff]=T[\ff] -V[\ff]$. Let us focus first on the shift of the ground state energy.  We find,  at leading order,  
\begin{align}
	\delta E_0&\approx\frac{1}{N}\qty[ \Th^{(2)} - \Vh^{(2)} -\frac12 \Th^{(0)}] .
\end{align}
For the gradient energy of the COM  wavefunction, we find (again at leading order)
\begin{equation}\begin{split}
    \hat{E}_2             &\approx \qty[\mathcal{K}_1^2 -6\mathcal{K}_2 ] \qty[\Th^{(2)}-\Vh^{(2)}]\\
&\quad- \mathcal{K}_1\qty[\Th^{(0)} - 2 \Vh^{(0)}]~.
\end{split}\end{equation}
As emphasized  in  the main  text  this  is the mean  result of our  work  and  demonstrates that quantum fluctuations of  the  COM  can lower the energy of a CCS state.  

%
\subsection{Cancellations Due to Normalization Conditions  \label{cancellations}}

In the previous section we found that terms such as $\langle \ff_0^{(1)} , \ff_0^{(1)}\rangle$ were absent at $\order{1/N}$ , and likewise terms such  as $\langle \ff^{(2)}_0, \ff^{(2)}_0 \rangle$ were absent at $\order{1/N^2}$.  In this section we outline that this is not an accidental cancellation, but is a direct consequence of the normalization conditions \cref{normalization-condition1,normalization-condition2,normalization-condition3}. 

To derive \cref{normalization-condition1,normalization-condition2,normalization-condition3} we demand that $\braket{\ff; N}{\ff ; N}=1$, and that this normalization is maintained order-by-order in $1/N$.   The exact expression for the overlap is given by 
\begin{equation}
	\braket{\ff}{\ff} = \int  \dd x \dd y  \ff\argl \ff^*\argr O(y; N)~. 
\end{equation}

At leading order, using  the delta-expansion of $O(y;  N)$  this is equivalent to demanding that 
\begin{equation}
	\langle\ff_0,  \ff_0 \rangle := \int_{-\pi}^\pi \ff^* _0(x) \ff_0(x)\dd  x= 1 ~,
\end{equation}
which is \cref{normalization-condition1}. At $\order{1/N}$ we find instead   
\begin{equation}
	\braket{\ff}{\ff}=  \langle \ff_0  ,\ff_0\rangle + \frac1N\qty[  2\mathrm{Re}\langle \ff_0 ,\ff_1\rangle  - \mathcal{K}_1 \langle \ff_0^{(1)}, \ff_0^{(1)}\rangle ].
\end{equation}
By requiring that this correction at $\order{1/N}$ vanish we arrive at \cref{normalization-condition2}.  Similarly, at  $\order{1/N^2}$  we have 
\begin{widetext}
\begin{equation}
	\braket{\ff}{\ff}=  \langle \ff_0  ,\ff_0\rangle + \frac1N[ ( 2\mathrm{Re}\langle \ff_0 ,\ff_1\rangle)  - \mathcal{K}_1 \langle \ff_0^{(1)}, \ff_0^{(1)}\rangle ] + \frac1{N^2}[ \mathcal{K}_2  \langle  \ff_0^{(2)},  \ff_0^{(2)}\rangle  - \mathcal{K}_1  ( 2\mathrm{Re}\langle  \ff_0^{(1)},\ff_1^{(1)} \rangle)  + 2 \mathrm{Re}\langle \ff_0 ,\ff_2\rangle + \langle \ff_1 ,\ff_1 \rangle ]~.
 \end{equation}
\end{widetext}
Our third normalization condition, \cref{normalization-condition3}, then follows from the requirement that the bracketed term of $\order{1/N^2}$ must  vanish.

Importantly, this exact same combination of terms is guaranteed to appear in our calculations  of  $E[\ff]$.    This  is most  clearly   illustrated   at  $\order{1/N}$. Let  us consider  just the term  
\begin{equation}\begin{split}
	&\int \dd x \dd y ~\mathcal{K}_1 \delta^{(2)}(y)\ff_0  \ff_0^*  \Vh(y)~\\
	&=\mathcal{K}_1 \Vh^{(2)}  - \mathcal{K}_1 \Vh^{(0)}  \langle \ff_0^{(1)} ,\ff_0^{(1)} \rangle ~.
\end{split}\end{equation}
Notice that when  the derivatives act  on the function $\ff_0$ it gives the same result as the normalization condition, but with an overall prefactor of $\Vh^{(0)}$. The same prefactor will also appear in the term 
\begin{equation}\begin{split}
&\int \dd x \dd y~ \delta(y)\qty[\ff_0(\alpha)  \ff_1^*(\beta) +   \ff_1(\alpha)  \ff_0^*(\beta) ]\Vh(y)\\
&=\Vh^{(0)} ( 2\mathrm{Re} \langle \ff_0 ,\ff_1\rangle)~ ,
\end{split}\end{equation}
where we have used $\alpha=x-\tfrac12 y$, and  $\beta=x+\tfrac12y$ for  shorthand.   Upon addition  of these two terms, we will have the combination that  corresponds to \cref{normalization-condition2}. This happens when \emph{all} of the derivatives  from the delta-expansion act on $\ff_0$;  this leaves no  derivatives left-over to act on $\Vh(y)$  and this ensures that the prefactor appearing in front of $\mathcal{K}_n \langle \ff_0^{(n)} ,\ff_0^{(n)}\rangle$ is $\Vh^{(0)}$. This is  why the gradient corrections to the COM wavefunction's energy appear at $\order{1/N^2}$ as opposed to $\order{1/N}$ as may be naively expected. The  same  cancellation  occurs  at $\order{1/N^2}$ but precludes terms of the form  $\langle \ff^{(2)}_0, \ff^{(2)}_0 \rangle$. 

%
\section{Many-body overlap functions \label{overlap-appendix}}
In the body of the main text we claimed that the functions $O(y; \aleph)$ could be expanded in the large $\aleph$ limit in a ``delta-expansion''
\begin{align}\label{overlap-expansion-app}
O(y;\aleph) &= \frac{1}{\abs{C(\aleph,\chi)}^2}\qty[\delta(y)+ \sum_{p>0} \frac{\mathcal{K}_p(\chi)}{\aleph^p} \delta^{(2p)} (y) ]~.
\end{align}
In this section we will justify this claim by making use of the properties of the Hartree wavefunctions $\psi_H(\theta)$. The results obtained in this section will allow us to obtain explicit expressions for $\mathcal{K}_1$,  and $\mathcal{K}_2$ in \cref{small-chi}.  As noted before, the $\aleph$-body overlap can be re-written as an exponentiated overlap of the Hartree states 
\begin{equation}
	O(y;\aleph)= \qty[~\braket{\psi_H ; ~x-\tfrac12 y}{\psi_H;~x+\tfrac12 y}~]^\aleph, 
\end{equation}
where
\begin{equation}\begin{split}
	&\braket{\psi_H ; ~x-\tfrac12 y}{\psi_H;~x+\tfrac12 y}\\
	&= \int \dd\theta \psi_H^*(\theta-[x-\tfrac12 y] ) \psi_H(\theta-[x+\tfrac12 y] )\\
	& = \int \dd \theta \psi_H(\theta+\tfrac12y) \psi_H(\theta-\tfrac12y) ~,
\end{split}\end{equation}
and, where we have used the fact that $\psi_H(\theta)$ is real. The form of the Hartree wavefunctions are known: they are given by appropriately scaled and shifted  Mathieu functions, with an auxiliary parameter $q(\chi)$ that can be determined exactly
\begin{equation}
	\psi_H(\theta)= \frac{1}{\sqrt{\pi} }\mathrm{ce}_0\qty (\tfrac12(\theta-\pi) ; q(\chi) ~)~.
\end{equation}
Thus, we have 
\begin{equation}\begin{split}
	&\braket{\psi_H ; ~x-\tfrac12 y}{\psi_H;~x+\tfrac12 y}\\
	&= \frac1\pi \int \dd\theta~ 
	\mathrm{ce}_0(\tfrac12 \theta; q) \mathrm{ce}_0(\tfrac12(\theta+y); q) ~.
\end{split}\end{equation}
Now for $\chi \ll 1$ we have that $q\sim 1/\chi^2$ such that $q$ is very large. In this regime the Mathieu functions are well approximated by parabolic cylinder functions, $D_n$, via Sips' expansion \cite{Olver2010} 
\begin{equation}\label{sips}
	\mathrm{ce}_0(z;q)\sim  C_0(q)\qty[ U_0(\xi; q) + V_0(\xi; q)]
\end{equation}
%
%
\begin{align}
	C_0(q)&\sim\qty[\frac{\pi \sqrt{q}}{2}]^{1/4}\qty[ 1 + \frac{1}{8\sqrt{q}} ]^{-1/2}\\
	 U_0(\xi~;q)&\sim D_0(\xi)- \frac{1}{4\sqrt{q}} D_{4}(\xi)\\
	  V_0(\xi~;q)&\sim - \frac{1}{16\sqrt{q}} D_{2}(\xi)\\
\end{align}
such that 
\begin{equation}
  \mathrm{ce}_0(z;~q)\sim \qty[ \frac{ \pi\sqrt{q} }{2} ]^{1/4} D_0(\xi)   + \order{\frac{1}{\sqrt{q}} }
\end{equation}
Introducing the variables  $\zeta=2 q^{1/4} \sin\tfrac{\theta}{2}$ we then find
\begin{equation}\begin{split}
  \psi_H(\theta;\chi)&\sim \qty[\frac{\sqrt{q}}{2\pi}]^{1/4} D_0(\zeta) + \order{\frac{1}{\sqrt{q}} }\\
  &= \qty[\frac{q}{(2\pi)^2}]^{1/8}\e^{-\sqrt{q} \sin^2\tfrac\theta2}+ \order{\frac{1}{\sqrt{q}} }.
\end{split}\end{equation}
Using the leading order behaviour for $\psi_H$,  the overlap can be expressed as a Bessel function
\begin{equation}\begin{split}
    &\braket{\psi_H ; ~x-\tfrac12 y}{\psi_H;~x+\tfrac12 y}\\
    &=\int_0^{2\pi} \dd \theta \psi_H^*(\theta-\tfrac12 y;\chi) \psi_H(\theta+\tfrac12 y;\chi) \\
    &\sim  \qty[\frac{q}{(2\pi)^2}]^{1/4}\int_0^{2\pi}  \dd \theta \e^{-\sqrt{q}\sin^2\tfrac{x-y/2}{2}} \e^{-\sqrt{q}\sin^2\tfrac{x+y/2}{2}} \\
    &=  \qty[\frac{q}{(2\pi)^2}]^{1/4}\ \int_0^{2\pi } \dd\theta \e^{\sqrt{q}\qty(1-\cos x \cos\tfrac{y}{2}) }\\
    &= \frac{I_0(\sqrt{q}\cos\tfrac{y}{2})}{\sqrt{2\pi q^{1/2}} \e^{\sqrt{q}}}
\end{split}\end{equation}
where $I_0(z)$ is the modified Bessel function of the first kind \cite{Olver2010}.  At the same order of accuracy we can instead write 
\begin{equation}\begin{split}
 \braket{\psi_H ; ~x-\tfrac12 y}{\psi_H;~x+\tfrac12 y}&\sim \frac{I_0(\sqrt{q}\cos\tfrac{y}{2})}{I_0(\sqrt{q})}~,
\end{split}\end{equation}
which is exact for $y=0$.  For most values of $y$ we can use a large-argument expansion for the Bessel function $I_0(z)\sim \e^{z}/\sqrt{2\pi z}$. For values of $y$ such that $\sqrt{q}\cos \tfrac{y}{2}\sim\order{1}$ it follows that $I_0(y) \sim \order{1}$ and so the overlap is $\order{q^{1/4}\e^{-\sqrt{q}}}$.   

When considering integrals on the interval $y\in[-\pi,\pi]$ it is therefore justifiable to neglect contributions from this exponentially suppressed region. Then, on the remainder of the interval, we can use the large-argument expansion of the Bessel function as a global approximation. This allows us to re-write the overlap as 
\begin{equation}\begin{split}
 &\braket{\psi_H ; ~x-\tfrac12 y}{\psi_H;~x+\tfrac12 y}\\
 &\sim \exp\qty[ 2\sqrt{q} \sin^2\tfrac{y}{4} - \tfrac12 \log \cos\tfrac{y}{2} + \order{\tfrac{1}{\sqrt{q}} } ] 
\end{split}\end{equation}
By extension the $\aleph$-body overlap assumes the form 
\begin{equation}\begin{split}
 O(y;\aleph)\sim\exp\left\{\aleph\qty[2\sqrt{q} \sin^2\tfrac{y}{4} - \tfrac12 \log \cos\tfrac{y}{2}] \right\}~,
\end{split}
\end{equation}
where we have neglected terms of $\order{1/\sqrt{q}}$ or smaller.  Trading $q$ for $\chi$ via $q\sim 4\chi^{-2}(1- \chi/4)$, we find  at the same order of accuracy
\begin{equation}
 O(y;\aleph)\sim \e^{\aleph\qty[\qty(\tfrac{4}{\chi}-\tfrac12) \sin^2\tfrac{y}{4} - \tfrac12 \log \cos\tfrac{y}{2} ]}~.
\end{equation}

%



\section{Small $\chi$ expansions \label{small-chi}} 
As noted in the main text,  $\Th(y)$'s leading order behavior as a function of $\chi$ is important.  We would like to  compute $\Th^{(0)}$ and $\Th^{(2)}$ and we will make use of  Sips' expansion for the ground state wavefunctions \cref{sips}
\begin{equation}
	\psi_H(x) \sim \qty[\frac{\sqrt{q}}{2\pi}]^{1/4}\qty[  D_0(\zeta) -\frac{1}{16\sqrt{q}}\mathfrak{D}(\zeta) ]~,
\end{equation}
where $\zeta=2q^{1/4}\sin\tfrac\theta2$, and 
\begin{equation}
	\mathfrak{D}(\zeta)= D_0(\zeta) + D_2(\zeta) + \frac14 D_4(\zeta)~.
\end{equation}
We are interested in  
\begin{equation}\label{Th-x}
	\Th(y)=\frac{ \chi^2}{2} \int \dd x \qty[\dv{x}\psi_H(x-\tfrac12 y)] \qty[\dv{x} \psi_H(x+\tfrac12 y)]~.
\end{equation}
It will be useful to have the following identities 
\begin{align}
   \dv{\zeta_\pm}{x} &=  q^{1/4} \cos\qty(\frac{x\pm \tfrac12 y}{2})= q^{1/4}  \qty(1-\frac{\zeta_\pm^2}{4\sqrt{q}} )^{1/2}\\
    \dv[2]{\zeta_\pm}{x}&= -\frac{q^{1/4}}{2}\qty(\frac{x\pm \tfrac12 y}{2})=-\frac{q^{1/4}}{4} \zeta_\pm~,
\end{align}
where $\zeta_\pm = \zeta(x\pm \tfrac12 y)$ with which we can  re-express \cref{Th-x} as 
\begin{equation}\begin{split}\label{Th-zeta}
	\Th(y)= -\frac{ \chi^2}{2}\sqrt{q}&\int \frac{\dd \zeta}{q^{\tfrac14}\sqrt{1-\frac{\zeta^2}{4\sqrt{q}}} }  \qty(1-\frac{\zeta_-^2}{4\sqrt{q}} )^{1/2}\\
	&\times
	\qty(1-\frac{\zeta_+^2}{4\sqrt{q}} )^{1/2}\psi_H'(\zeta_-)\psi_H'(\zeta_+)~.
\end{split}\end{equation}
At leading order in $1/\sqrt{q}$ we have 
\begin{equation}\label{Th-zeta-LO}
	\Th(y)\sim -\frac{ \chi^2}{2}\sqrt{q}\int\frac{ \dd \zeta}{\sqrt{2\pi}} D_0'(\zeta_-)D_0'(\zeta_+)~.
\end{equation}
This leads immediately to the result 
\begin{equation}\label{Th0-LO}\begin{split}
	\Th^{(0)}&\sim -\frac{ \chi^2}{2}\sqrt{q}\int\frac{ \dd \zeta}{\sqrt{2\pi}} D_0'(\zeta)D_0'(\zeta)\\
	&=-\frac{\sqrt{q}}{8}\chi^2 ~. 
\end{split}\end{equation}
Next, to calculate $\Th^{(2)}$ we must act with $\dv[2]{y}$ on \cref{Th-zeta-LO}.  A useful identity is
\begin{equation}
\begin{split}
\dv[2]{y}&\qty[f(\zeta_-)g(\zeta_+) + f(\zeta_+)g(\zeta_-)] \\
=~& \qty[\dv[2]{\zeta_+}{y}+ \dv[2]{\zeta_-}{y}]\qty[f'g + g' f]\\
&+ 4 \dv{\zeta_+}{y} \dv{\zeta_-}{y} f' g'\\
&  + \qty[\qty(\dv{\zeta_+}{y})^2\qty(\dv{\zeta_-}{y})^2][f''g + g''f]~.
\end{split}
\end{equation}
which, holds when $y=0$.  We can insert this identity underneath the integral after acting with the derivative operator. This will give us an integral representation for $\Th^{(2)}:= \Th''(y=0)$.  Using the explicit forms of the derivatives, 
\begin{align}
	\dv{\zeta_\pm}{y} &= 
	\pm \frac{q^{1/4} }{2}\qty(1-\frac{\zeta_\pm^2}{4\sqrt{q}} )^{1/2}\\
\dv[2]{\zeta_\pm}{y}&=  
-\frac{\zeta}{16}~,
\end{align}

we find

\begin{equation}\begin{split}
		\Th^{(2)}&\sim -q \frac{ \chi^2}{2} \int\frac{ \dd \zeta}{\sqrt{2\pi}} 2 \qty[D'_0 D'_0 - 2 D_0'''D_0']\\
		&= 2q  \chi^2 \int\frac{ \dd \zeta}{\sqrt{2\pi}} D''_0 D''_0   ~\\
		&= \frac{3 q\chi^2}{32} .
\end{split}\end{equation}

where we have used the leading order approximation for  $ \dd \zeta_\pm/ \dd y$ and neglected the contribution from terms proportional to $\dd^2\zeta_\pm/\dd^2 y$ because they are subleading. To obtain the second equality we integrated by parts, however at higher orders in $1/\sqrt{q}$ one needs to be careful to keep track of factors of $\zeta^2$ in the integrand. 

When calculating $\Vh^{(0)}$   and  $\Vh^{(2)}$ we need to work  beyond leading order, because the leading order piece cancels in \cref{e2-eval}.  We are interested in 
\begin{equation}\begin{split}
\Vh(y)=\frac12  \int &\dd x_1 \dd x_2 \psi_H(x_1+\tfrac{y}{2}) \psi_H(x_1-\tfrac{y}{2}) \\
&\times \psi_H(x_2+\tfrac{y}{2}) \psi_H(x_2-\tfrac{y}{2})\cos(x_1-x_2)
\end{split}
\end{equation}
which can be  re-written  as 
\begin{align}
	\Vh(y)&= \frac12\qty[I_C(y)^2 + I_S(y)^2]\\
	I_C(y)&= \int \dd x \psi_H(x+\tfrac{y}{2}) \psi_H(x-\tfrac{y}{2}) \cos(x)\\
		I_S(y)&= \int \dd x \psi_H(x+\tfrac{y}{2}) \psi_H(x-\tfrac{y}{2}) \sin(x)
\end{align}
Importantly $I_S(0)=0$, $I'_S(0)=0$,  and $I'_C(0)$  such that 
\begin{equation}
	\Vh^{(0)}=\frac12 I_C^2(0) \qq{and} \Vh^{(2)}= I_C(0)I_C''(0)~,
\end{equation}
so we can focus exclusively on the integral $I_C(y)$.  Re-writing this in terms of $\zeta$ and keeping only terms to order  $1/\sqrt{q}$ we arrive at 
\begin{widetext}
\begin{equation}\label{IC-NLO} 
\begin{split}
	I_C(y)=\frac{1}{\sqrt{2\pi}} &\bigg[ \int D_0(\zeta_-)D_0(\zeta_+)\qty(1-\frac{3\zeta^2}{8 \sqrt{q}} ) \dd \zeta - \frac{1}{16\sqrt{q}} \int D_0(\zeta_-)\mathfrak{D}(\zeta_+)+D_0(\zeta_+)\mathfrak{D}(\zeta_-)  \dd \zeta \bigg]  
\end{split}
\end{equation}
Evaluating at $y=0$ sets $\zeta_\pm=\zeta$ and we find 
\begin{equation}
\begin{split}
	I_C(0)&=\frac{1}{\sqrt{2\pi}}\bigg[ \int D_0(\zeta)D_0(\zeta)\qty(1-\frac{3\zeta^2}{8 \sqrt{q}} ) \dd \zeta-\frac{1}{8\sqrt{q}} \int D_0(\zeta)\mathfrak{D}(\zeta) \dd \zeta \bigg]
	=1-\frac{1}{2\sqrt{q}}
\end{split}
\end{equation}
\end{widetext}

To find $I_C''(0)$ we must act on \cref{IC-NLO} with $\dv[2]{y}$. Being careful to retain sub-leading terms we find 
\begin{equation}
\begin{split}
	I_C''(0)&= 	-\frac{\sqrt{q}}{\sqrt{2\pi}} \bigg[\int \frac{D'_0D'_0- D''_0D_0}{2} \qty(1-\frac{5\zeta^2}{8 \sqrt{q}} ) \dd \zeta \\
	&~~~~~~~~+\frac{2}{\sqrt{2\pi}} \int D_0'D_0 \frac{\zeta}{16} \dd \zeta - \frac{1}{8\sqrt{q}} \int D'_0\mathfrak{D}' \dd \zeta \bigg]  \\
	&= -\frac{\sqrt{q}}{4}\qty(1-\frac3{4\sqrt{q}})
\end{split}
\end{equation}

Using $\Vh^{(0)}=\tfrac12 [I_C(0)]^2$, $\Vh^{(2)}=I_C''(0)I_C(0)$, and the small-$\chi$ behavior of $q$ \cite{Plestid2018a},  
\begin{equation}
q\sim\frac{4}{\chi^2}\qty[1-\frac{\chi}{4}+\order{\chi^2}]
\end{equation}
we then find
\begin{align}
	\Vh^{(0)}&=\frac{1}{2}\qty[1-\frac{1}{ \sqrt{q}}  + \order{\frac{1}{q} } ]\\
	\nonumber &=\frac{1}{2}  \qty[1-\frac\chi2+\order{\chi^2}]\\
	\vspace{6pt}
	\Vh^{(2)}&= -\frac{\sqrt{q}}{4}\qty[1 -\frac{5}{16 \sqrt{q}} + \order{\frac{1}{q} }]\\
	\nonumber &=-\frac{1}{2\chi}\qty[1 -\frac{3\chi}{4}  +\order{\chi^2}] \\
	\vspace{6pt}
	\Th^{(0)}&= \chi^2\times \frac{\sqrt{q}}{8}\qty[1 +  \order{\frac{1}{\sqrt{q}} } ]\\
	\nonumber &= \frac{\chi}{4}\qty[1+\order{\chi}] \\
	\vspace{6pt}
	\Th^{(2)}&= \chi^2\times\qty(- \frac{3q}{32})\qty[1 +  \order{\frac{1}{\sqrt{q}} } ]\\
	\nonumber &= -\frac{3}{8} \qty[1+ \order{\chi}]~.
\end{align}

\bibliography{hmf-dyn.bib}

\end{document}